\documentclass[traditabstract]{aa} 
\usepackage{graphicx}
\usepackage{txfonts}
\usepackage{color}
\begin{document}
   \title{The Kepler and Hale observations of V523 Lyr}

   \subtitle{}

   \author{E. Mason
          \inst{1}
          \and
          S.B. Howell\inst{2}\fnmsep\thanks{Visiting Astronomer, Mt. Palomar Observatory}
          }

   \institute{INAF-OATS, Via G.B. Tiepolo 11, 34143, Trieste, IT\\
              \email{emason@oats.inaf.it}
         \and
             NASA Ames Research Center,
Moffett Field, CA 94035, USA\\
             \email{steve.b.howell@nasa.gov}
             }

   \date{Received: ; accepted:}

 
  \abstract
   {We present new observations of the cataclysmic variable V523~Lyr, a member of the open cluster NGC 6791.
The Kepler Space telescope obtained photometric observations of this source and we examine the nearly 3 year long light curve. The observations show numerous small amplitude outbursts recurring on average every 33 days, intermittent quasi-periodic oscillations, and a significant fully coherent period of $\sim$3.8 hr which we identify as the orbital period of the binary.
Contemporaneous optical spectroscopy of V523~Lyr reveals a faint blue source with broad Balmer absorption lines containing narrow emission cores. H$\alpha$ is in emission above the continuum. The low amplitude of the photometric signal and no detected velocity motion suggest a low orbital inclination. We discuss the properties of V523~Lyr and show that it is a member of the growing group of anomalous Z Cam type CVs, systems which show stunted outbursts, light curve standstills, and occasional deep drops in brightness.
   }


   \maketitle
%

\section{Introduction}

The original Kepler mission was dedicated to the search and discovery of extra-solar planets via transit detection and particularly focused on determining the frequency of earth-size planets orbiting solar-type stars (Borucki et al. 2010). However, through its Guest Observer program, Kepler also targeted a number of other objects including variables stars such as cataclysmic variables (CVs). These 
objects are interacting binary systems in which a white dwarf (WD) accretes matter from a low-mass main sequence companion, either through magnetically controlled accretion (magnetic CV) or through the formation of an accretion disk surrounding the WD. Kepler's high photometric precision and prolonged light curve observations of CVs are proving invaluable in the understanding of their variability phenomena, in particular the outbursts produced by disk instabilities and semi-periodic variations (e.g. Wood et al. 2011, Cannizzo et al. 2012, Osaki \& Kato 2013, Still et al. 2010). Nine CVs were the known prior to launch and targeted for observation by the Kepler mission (Howell et al. 2013) and over ten additional CVs have been discovered in the Kepler field (e.g. Scaringi et al. 2013). Among the known CVs, V523~Lyr was especially interesting since it belongs to the open cluster NGC~6791 and the faint star is poorly classified/characterized. CVs in star clusters might form via different mechanisms than those 
in the field 
and are valuable probes of cluster evolution and dynamics. However, while predicted, known CVs in open clusters are rare (up to year 2015 only 4 have been confirmed, among which 3 are in NGC~6791; van den Berg et al. 2013, Williams et al. 2013). Here we present the V523 Lyr light curve obtained by Kepler as well as accompanying Mt. Palomar 200" Hale spectroscopy.

\section{V523 Lyr observational history}
NGC~6791 is an old ($\sim$8 Gyr) metal rich ([Fe/H]$\geq$+0.4) open cluster about 4.3 kpc away from our Sun (e.g. Chaboyer et al. 1999, Carraro et al. 2006, Gratton et al. 2006, Grundahl et al. 2008; and references therein). The cluster has been the object of a number of variable star surveys due to its relatively large number of blue stragglers and ground-based searches for open cluster exoplanets. As a result, V523 Lyr has a somewhat extended set of monitoring observations even preceding the nearly three years of Kepler observations. 

During the first ``CCD-era" survey of the cluster by Kaluzny \& Udalski (1992), V523~Lyr ($\equiv$B7) was discovered as a blue star (V=18.329 mag, B-V=0.139, and V-I=0.336 mag). It was subsequently identified as a cluster variable ($\equiv$V15, V$_{max}$=17.67 mag, B-V=0.092 mag) by  Kaluzny \& Rucinski (1993) who observed a progressive fading of 0.17 mag across their 4 night observing run. Combining their new and previous observations, they suggested that V523~Lyr was a (detached) binary hosting a red dwarf and a hot subdwarf. V523~Lyr was first suggested to be an interacting binary by Liebert et al. (1993) who obtained optical spectroscopy (3650-5225 \AA). Liebert et al. reported a relatively flat spectral energy distribution with shallow H-Balmer and  HeI absorptions containing narrow emission cores. 

Kaluzny et al. (1997) monitored V523~Lyr for 30 nights between Sep 8 and Nov 1 1996. Their V and I band light curves showed that V523~Lyr decreased by 3(2) mag in V(I) in less than 10 days and returned back to V$\sim$18(I$\sim$17.8) mag in $\sim$25 days (the entire rise time, however, was not covered by the monitoring). Their light curves covered only one small outburst of amplitude $\leq$0.6 (0.4) mag in V(I). Kaluzny et al. (1997) complemented their photometric observations with two spectra taken in early 1997 when V523~Lyr was at V=18.13 mag and V$\sim$19 mag. The brighter spectrum showed a blue continuum, H$\alpha$ in emission with the remaining Balmer lines in absorption, possibly including weak emission cores. The photometric time series obtained during the same night (April 1997) revealed no flickering or orbital modulation. The fainter spectrum, taken on May 1997, revealed only weak H absorptions. 

In July 2001, V523 Lyr was observed again in some detail as part of the PISCES survey (Mochejska et al., 2002). Their 18 non-continuous nights of photometric monitoring detected a 0.6 mag outburst of V523~Lyr. The outburst was characterized by a slow rise ($\sim$4-5 days) and a similarly slow decline ($\sim$5-7 days). The post-outburst magnitude was $\sim$0.2 mag fainter than pre-outburst and returned to normal after about 2-3 days. Mochejska et al. (2003) obtained long-term monitoring of the cluster and collected 123 nights of observation across almost 6 years with the aim of properly characterizing variables in NGC~6791. Within this monitoring, V523~Lyr showed a second deep fading by 3 mag and 4 outbursts of 0.5-1.0 mag in V. Mochejska et al. (2003) estimated an outburst recurrence time of 25.4 days.  These authors realized that both the outburst recurrence time and outburst amplitude were compatible with those observed in Z Cam type CVs, while the 3 mag drops in light were reminiscent of VY Scl 
class. Hence, they remained uncertain about the specific CV class to which V523~Lyr belongs. 

De Marchi et al. (2007) added 10 more days of photometric monitoring of NGC~6791 during which time V523~Lyr was found to display fairly erratic photometric behavior showing 0.06 mag intra-night variations and up to 0.15 mag differences across nights. 

At non-optical wavelengths, a Chandra X-ray survey of the cluster (van den Berg et al., 2013)
detected V523~Lyr at L$_X$=6.4E30 erg/s (assuming a cluster distance of 4.1 kpc and a kT=2eV optically thin plasma). van den Berg et al. comment that the X-ray colors and luminosities of the CVs in their survey match typical values expected for this class of object. We shall add, as a comparison, that the dwarf nova SS Cyg has a measured X-ray flux of L$_X\sim$1E33 erg/s and 1E32 erg/s in quiescence and outburst, respectively (Fabbiano et al. 1978).
In the infrared, V523 Lyr remained undetected in the 2$^{nd}$ 2MASS data release (Hoard et al. 2002). 


\begin{table}
\label{KepLog}
 \centering
\caption{Kepler log of observation in Long Cadence (LC) mode. ``Start'' and ``end'' dates are BJD$-$2454833.
Quarters 9, 13, and 15 also had observations in SC mode (KIC ID 100003514, 100003517, 100004034 and 100004109). Kepler Long cadence data sets 100003514, 100003517 and 100003520 are identical, therefore only 100003514 is listed in the table.}

\begin{tabular}{cccc}
KIC ID & Quarter & START & END \\ 
100002727 & 6 & 539.4 & 629.3 \\ 
100003096 & 7 & 630.2 & 719.6 \\ 
100003411 & 8 & 735.3 & 802.4 \\ 
100003514 & 9 & 808.5 & 905.9 \\ 
100003562 & 10 & 906.8 & 1000.7 \\ 
100003608 & 11 & 1001.2 & 1098.3 \\ 
100003908 & 12 & 1099.3 & 1182.0 \\
100004034 & 13 & 1182.7 & 1273.1 \\ 
100004075 & 14 & 1274.1 & 1371.1 \\ 
100004109 & 15 & 1373.5 & 1471.1 \\ 
100004182 & 16 & 1472.1 & 1558.0 \\ 
 &  &  &  \\ 
\end{tabular}
\end{table}

\section{Kepler photometry}

   \begin{figure}
   \centering
   \includegraphics[height=7.5cm,angle=0]{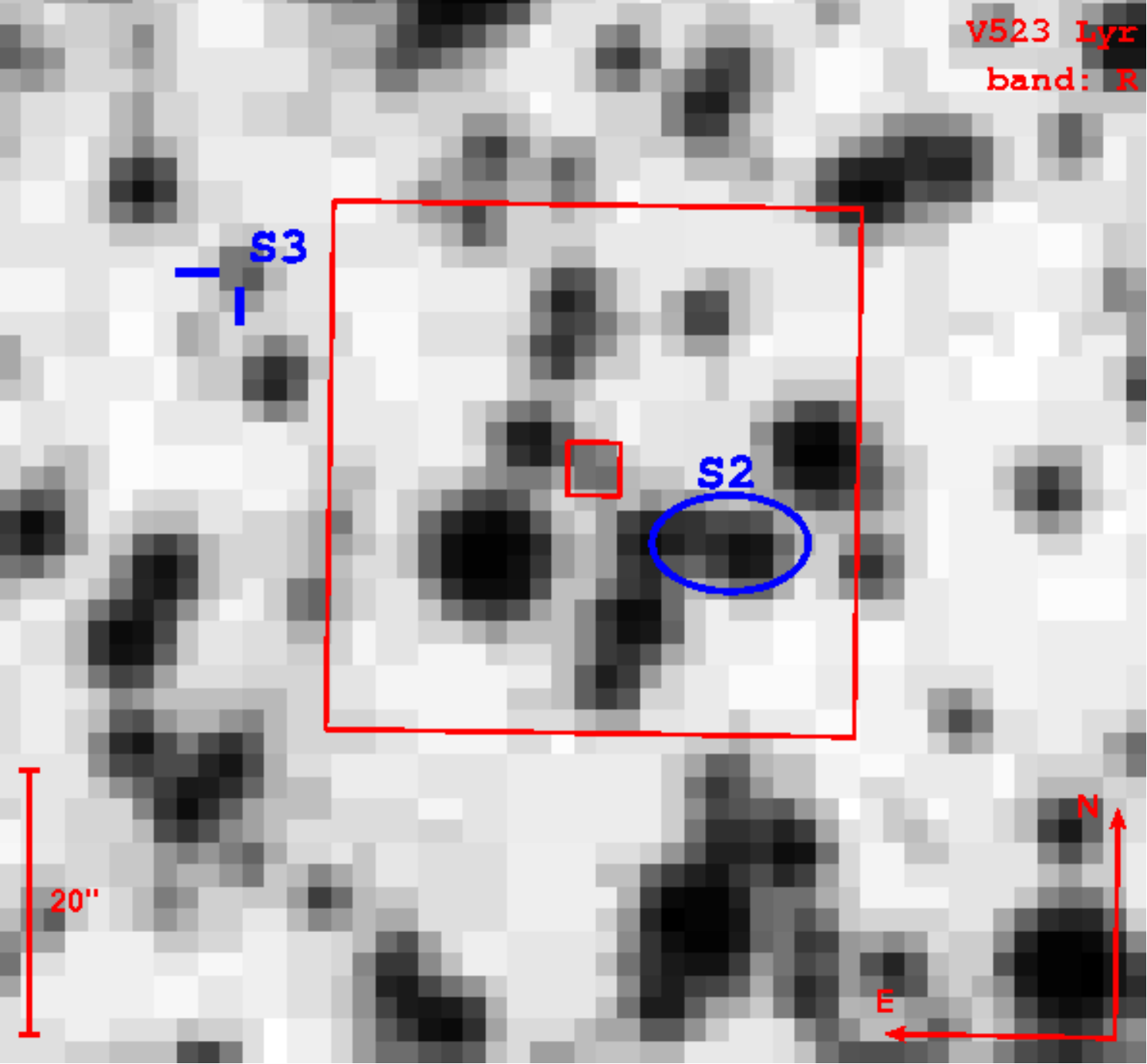}
   \caption{ SDSS image covering the Kepler field of view for V523 Lyr. The large red square indicates the typical aperture size ($\sim$10x10 pix$^2$, but often not rectangular in shape- see Kinemuchi et al., 2012) used for V523 Lyr in the standard Kepler pipeline to produce simple aperture photometry (SAP) light curves. The small red square indicates the Kepler pixel size (4'') and is shown here centered on the V523 Lyr position. Typical Kepler PSFs (noted in the bright stars) are 4 pixels across. S2 and S3 indicate the two confounding variable stars near V523 Lyr which occasionally fell into the V523 SAP aperture. See Appendix A for more details.}
              \label{fov}%
    \end{figure}

Kepler stared at V523 Lyr for 2.7 years (1019 days). Kepler data were down-loaded and archived in quarters ($\sim$3 months) between which the entire spacecraft, and thus focal plane, is rotated 90 degrees, placing targets on a new set of CCDs. After one year,
targets return to their original locations. V523 Lyr was observed for a total of 11 quarters (quarters 6 through 16) both in long
cadence (LC, 30 minute integration time) and short cadence (SC, 1 minute integration time). SC observations  covered 18 months (quarters 6, 13 and 15) but due to the object faintness, we discuss here only the LC data\footnote{A few SC data sets were analyzed similarly to the LC ones, as a cross check. They delivered consistent results, although with noisier periodograms due to the data quality and interval duration.}. 
Stars that were observed by Kepler are given a Kepler Input catalog (KIC) identification number (Brown et al. 2011). This number is fixed throughout the Kepler mission if the object was listed in the KIC. V523 Lyr was too faint to be in the catalog, thus it was assigned a special KIC number for each quarter of observation and for both its LC and SC data sets. The LC data sets we use in this paper are listed in Table~1. 

V523 Lyr is fairly faint source and located in a relatively crowded field. Fig.\ref{fov} shows the typical Kepler photometric aperture size (large square) used for V523~Lyr, the additional stars that would be contained in it, and the Kepler CCD pixel size. Clearly, the standard Kepler simple aperture photometry (SAP) light curve would include more than one object and different objects in each quarter due to the 90 degree roll. This might be irrelevant in the case of a bright target or with large photometric amplitude, but is not otherwise. In fact, the standard Kepler SAP light curve of V523 Lyr is biased by two other variable stars in the same field (see Appendix). Thus, we could not use the standard pipeline produced light curve and worked instead with the individual calibrated target pixel files, ({\sf targ.fits}) available for every source in the MAST archive. These files are small ``postage size" sets of pixels collected for each target for each cadence. For V523 Lyr, this means one image file for every 
30-min observation for the entire 2.7 years of data.
 
\begin{figure*}
\centering
\includegraphics[width=18cm,angle=0]{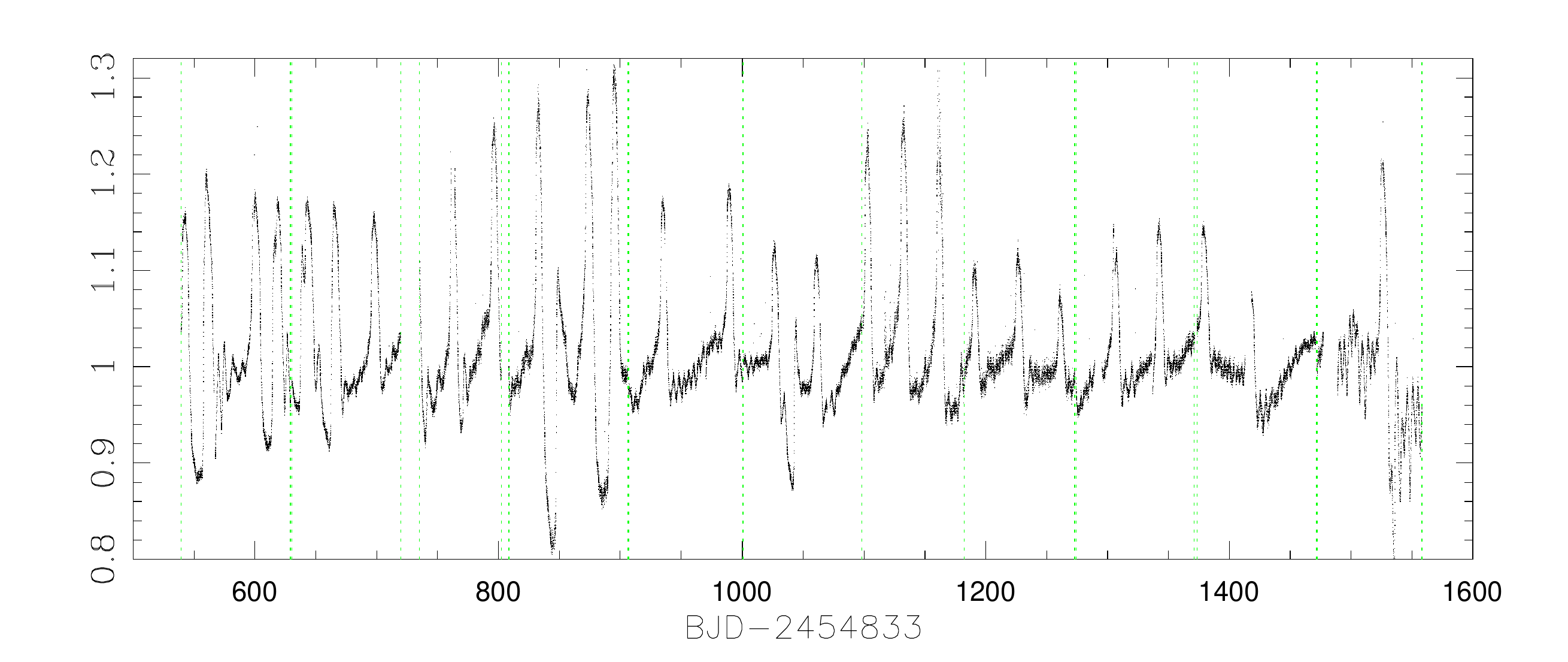}
\caption{The complete Kepler LC data set for V523 Lyr. Dashed green vertical lines mark the beginning and the end of each quarter. Y units correspond to median-normalized counts where each quarter has been normalized independently and stitched together. Note the numerous small outbursts with dips often following them, and the system becomes slightly fainter after each outburst and rises in brightness until the next. See text for more details.}
\label{1rawLC}%
\end{figure*}

For every {\sf targ} file, we examined each pixel's light curve to identify those pixels which contained the V523 Lyr signal alone. We then extracted the flux from those individual pixels and coadded the signal to produce each of our light curve points. The best results were obtained by selecting an aperture of 2 to 5 pixels in size (depending on the quarter), the majority being of 4 pixels (i.e., the Kepler PSF). Note that the inclusion of additional pixels containing the PSF wings usually added trends to the data and generally degraded the SNR. Details about each aperture size used and their pixel IDs are reported in Table~2. Each quarter's extracted light curve was then normalized by its global median value and all 2.7 years of data were stitched together. Our final light curve is shown in Fig.\ref{1rawLC}. 

\begin{table}
 \label{apsize}
 \caption{Quarter (QRT) aperture size (in pixels) and x,y pixel coordinates for each of the custom apertures adopted to produce the V523 Lyr LC light curve presented in this paper. Pixel coordinates (x,y) correspond to the CCD columns and rows (see FITS header keywords {\sc crval1p} and {\sc crval2p}). }
\centering
\scriptsize
\begin{tabular}{ccl}
QRT & SIZE & PIX COORDs (COL,RAW) \\
6 & 4 & (161,851) (161,852) (162,851) (162,852) \\ 
7 & 5 & (160,845) (161,844) (161,845) (161,846) (162,845) \\ 
8 & 5 & (156,844) (156,845) (157,844) (157,845) (157,846) \\ 
9 & 2 & (156,851) (157,851) \\ 
10 & 4 & (161,851) (161,852) (162,851) (162,852) \\ 
11 & 5 & (160,845) (161,845) (161,846) (162,845) (161,844) \\ 
12 & 4 & (156,844) (156,845) (157,844) (157,845) \\
13 & 3 & (156,850) (157,849) (157,850) \\ 
14 & 4 & (161,851) (161,852) (162,851) (162,852) \\ 
15 & 5 & (160,845) (161,844) (161,845) (161,846) (162,845) \\ 
16 & 4 & (156,844) (156,845) (157,844) (157,845) \\ 

\end{tabular}
\end{table}

The method used above to produce the light curve for V523~Lyr reduced the typical long-term trends sometimes observed for stars across quarters (Kinemuchi et al. 2012). As much as most of each quarters light curve matched end-to-end with the neighbor ones, we did not deem it necessary to apply any de-trending correction. While some long term trends are still apparent in Fig.2, we found that when we applied cotrending corrections, we could not always find a combination of the detrending vectors which would remove such long term trends in a satisfactory way (e.g. most applications increased the overall trends in parts of the light curve and generally increased the noise in the light curve).
With no application of a de-trending process, it may be that the exact shape of each quarter's light curve remains somewhat uncertain. Thus, one has to be careful when interpreting the data keeping in mind that 1) long term modulations can produce spurious low frequency signals in the periodogram analysis and 2) the exact shape and amplitude of each quarter's global and local light curve is somewhat uncertain, limiting the characterization of the outburst morphologies.

\section{Mt. Palomar 200" Hale spectroscopy}

V523 Lyr was observed with the Double Spectrograph (DBSP) mounted on the 200" Hale telescope located at the 
Mt. Palomar observatory. DBSP is a double beam, low to mid-resolution spectrograph capable of covering the wavelength range 3100-10000 \AA\, depending on the choice of grating. 
Dichroic filter D-55 was used to split light between the blue and red arms. The blue arm used a 1200 l mm$^{-1}$ grating providing R$\sim$7700 and covered 1500\AA\, of spectrum. The red arm used a 1200 l mm$^{-1}$ grating providing R$\sim$10,000 and covered only 670\AA. The slit width was set to 1$^{''}$ and the usual procedures of observing spectrophotometric stars and arc lamps were adhered to. Red spectra were wavelength calibrated with a HeNeAr lamp while the blue arm used a FeAr lamp. Data reduction was done using IRAF two- and one-dimensional image reduction routines for spectroscopic observations and we produced a final one-dimensional spectrum for each observation. 

The observations of V523~Lyr were carried out on UT July 1 and 2, 2013, taking 7 and 8 exposures per night, respectively, each 1200 sec long. The nights were clear and provided stable seeing near 1$^{''}$. We did not have contemporaneous photometry of V523~Lyr, but have estimated its brightness during our spectroscopic observations in two ways.
First, we estimated its brightness ``by eye"  using the slit viewer camera and comparing our target with other stars observed throughout the evening.  The method has large uncertainties of the order of 0.5-0.75 mag. We estimated V523 Lyr to be near V=20-21 mag.  A second more robust estimate was based on the raw counts collected through the slit for the standard star Feige 98 and V523 Lyr itself. This has an uncertainty due to slit losses, but the seeing was stable during the nights thus these are minimized. Based on a comparison of the counts received through the slit for both stars, adjusted for their respective integration times, we find that V 523 Lyr was near a B magnitude of 20.4 +/- 0.5 mags during the two nights. The magnitude of V523 Lyr during our observations was fainter than that reported in the literature for previous spectroscopy.

\section{Kepler light curve analysis}
Cataclysmic variables vary on a large range of time scales, from minutes to years. In particular, outbursts occur at a cadence of days/weeks to years while orbital periods range typically from 1.3 hr to $\sim$10 hr (with few exceptions of CVs showing period of the order of days).  Therefore, the Kepler monitoring allows us to probe both the outburst's recurrence time and light curve characteristics/morphology as well as search for the orbital period of the CV. 


\begin{table}
\label{otb}
\caption{Outburst parameters for V523~Lyr. Time is given in the same truncated units as in Fig.\ref{1rawLC}.
See text for more details.}
\centering
\begin{tabular}{ccccc}
ID & Time(max) & Peak Flux & FW0.1M  & FW0.15M  \\
& & & (days) & (days) \\
1 & 542.56 & 1.155 & 5.72 & NA   \\
2 & 560.35 & 1.195 & 6.29 & 7.42  \\
X(1)& 600.12 & 1.176 & NA & NA  \\
4 & 618.65 & 1.169 & 7.21 & 8.40  \\
5 & 643.07 & 1.170 & 9.83 & 11.07 \\
6 & 664.93 & 1.163 & 6.29 &  7.72 \\
7 & 698.20 & 1.153 & 6.68 & 11.85 \\
10 & 796.73 & 1.245 & 4.82 & 5.99 \\
11 & 832.59 & 1.270 & 4.17 & 4.90 \\
12 & 873.56 & 1.279 & 5.07 & 6.11 \\
13 & 895.14 & 1.304 & 4.39 & 5.52 \\
X(2) & 934.90 & 1.170 & NA & NA \\
15 & 989.14 & 1.183 & 5.31 & 9.40 \\
16 & 1026.79 & 1.120 & 6.64 & 11.24 \\
X(3) & 1061.22 & 1.125 & NA & NA \\
18 & 1102.90 & 1.235 & 4.52 & 5.58 \\
19 & 1132.69 & 1.250 & 4.80 & 5.93 \\
21 & 1190.69 & 1.100 & 14.92 & NA \\
22 & 1226.10 & 1.113 & 10.65 & NA \\
23 & 1260.94 & 1.071 & NA & NA \\
25 & 1342.37 & 1.147 & 5.97 & NA \\
26 & 1379.27 & 1.143 & 9.48 & NA \\
\end{tabular}
\end{table}

\subsection{Outbursts and dips}

\begin{figure*}
   \includegraphics[width=19cm]{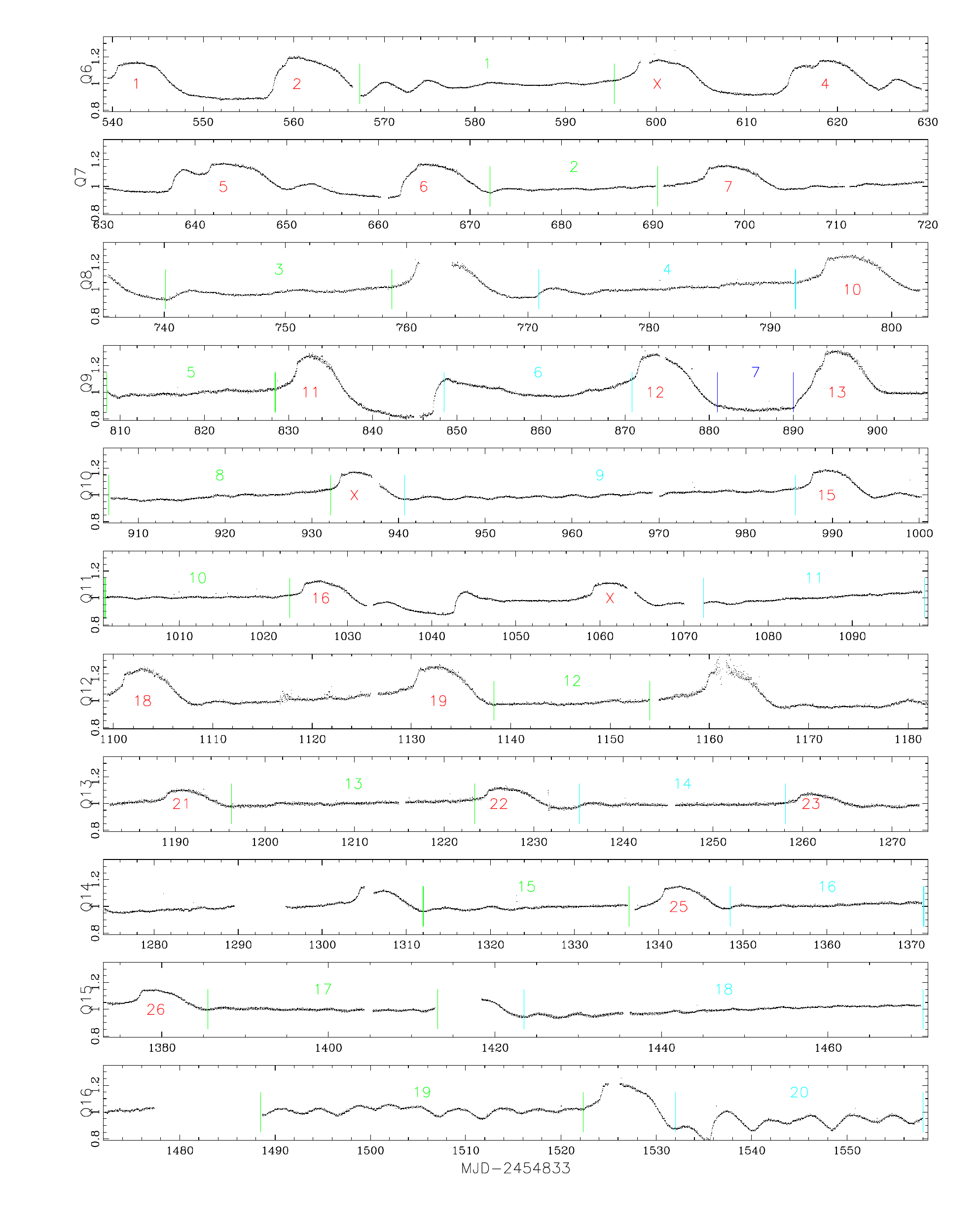}
   \caption{Kepler V523 Lyr light curve shown for each quarter. Pairs of vertical colored lines delimits quiescent intervals considered in the time series analysis for period searching. The intervals are also identified by a sequential number of matching color. Red numbers mark the outbursts considered for their characterization. The 3 outbursts marked with an X were considered only for the estimate of the recurrence time using their peak value.  See text for more details. }  
              \label{qrts}%
   \end{figure*}

The V523 Lyr light curve is quite complex as it shows outbursts, dips and oscillations. We list in Table~3 outburst parameters such as time of maximum, peak flux, and outburst duration. The parameters were measured on a 33 point running boxcar smoothed light curve. The duration of each outburst has been measured in two different ways: at a drop of 0.1 in normalized flux from maximum (FW0.1M) and at 0.15 in normalized flux from maximum (FW0.15M), whenever applicable. We observed 28 outbursts during the $\sim$1000 days of Kepler monitoring, but could measure just 19 since the remaining 9 either had too little amplitude or were ``incomplete'', i.e. have gaps effecting either the measure of the peak or the duration of the outburst. For three of the incomplete outbursts we could measure the maximum. All the outbursts listed in Table~3 are marked in Fig.\ref{qrts} with their sequential ID number. The outbursts for which it was possible to measure just the maximum time and peak intensity are identified 
by an ``X'' in the figure and the 
table. 

Outbursts of V523~Lyr have recurrence times of the order of weeks with a minimum recurrence time of $\sim$22 days and a maximum of $\sim$54 days (peak to peak). Mochejska et al. (2003) suggested 25.4 days as the recurrence time while our light curve finds 33$\pm$4 days as the mean value.
The outburst duration is of the order of $\sim$1 week (Table 3). 
The rise tends to be faster than the decline and most outbursts show a slow brightening before the final jump to maximum (see Fig. \ref{1rawLC} and \ref{qrts}). It is also evident that a few of the observed outbursts show the so called precursor (e.g. outburst \# 2, 4 and 5 and possibly also 6 and 12). However, these are only marginally longer and not necessarily the brightest. All the outbursts are within 10-30\% of the ``quiescent level''  (normalized at 1), implying brightenings of only $-$0.1-0.3 mag. 
The outburst amplitudes observed in the Kepler observations are quite smaller than the 0.6 and 0.4 mag  ``mini-outburst" shown by Kaluzny et al. (1997) or the 0.5-1.0 mag outbursts seen by Mochejska et al. (2003).  At least a small part of this discrepancy can be explained by the different transmission function of the Kepler bandpass\footnote{Close to a white-light filter (Koch et al. 2010).} compared to the Johnson filters used in the previous work. 


Either way, the observed outburst variations are just fractions of a magnitude and do not match typical dwarf novae outbursts (DN, of the U Gem or SU UMa type). They rather resemble the ``stunted" outbursts of some nova likes (NL), old novae and anomalous Z Cam stars (e.g. Honeycutt et al. 1998, Simonsen 2011).  Taken all together, the light curve properties of V523~Lyr are similar to another recently studied Kepler NL, KIC 9406652 (Gies et al., 2013).

The observed dips seen in the V523~Lyr light curve are also of small amplitude, amounting to $\sim$0.1-0.2 mag in the Kepler bandpass. Their duration is similar to the outburst, i.e. of the order of 1 week, and, if present, they always follow an outburst. In particular, they never appear isolated as previously reported for some NL or old novae observed by Honeycutt et al. (1998). The dips are not flat bottomed but typically display a relatively slow decline and a faster rise. 

\begin{figure}
   \includegraphics[width=9cm]{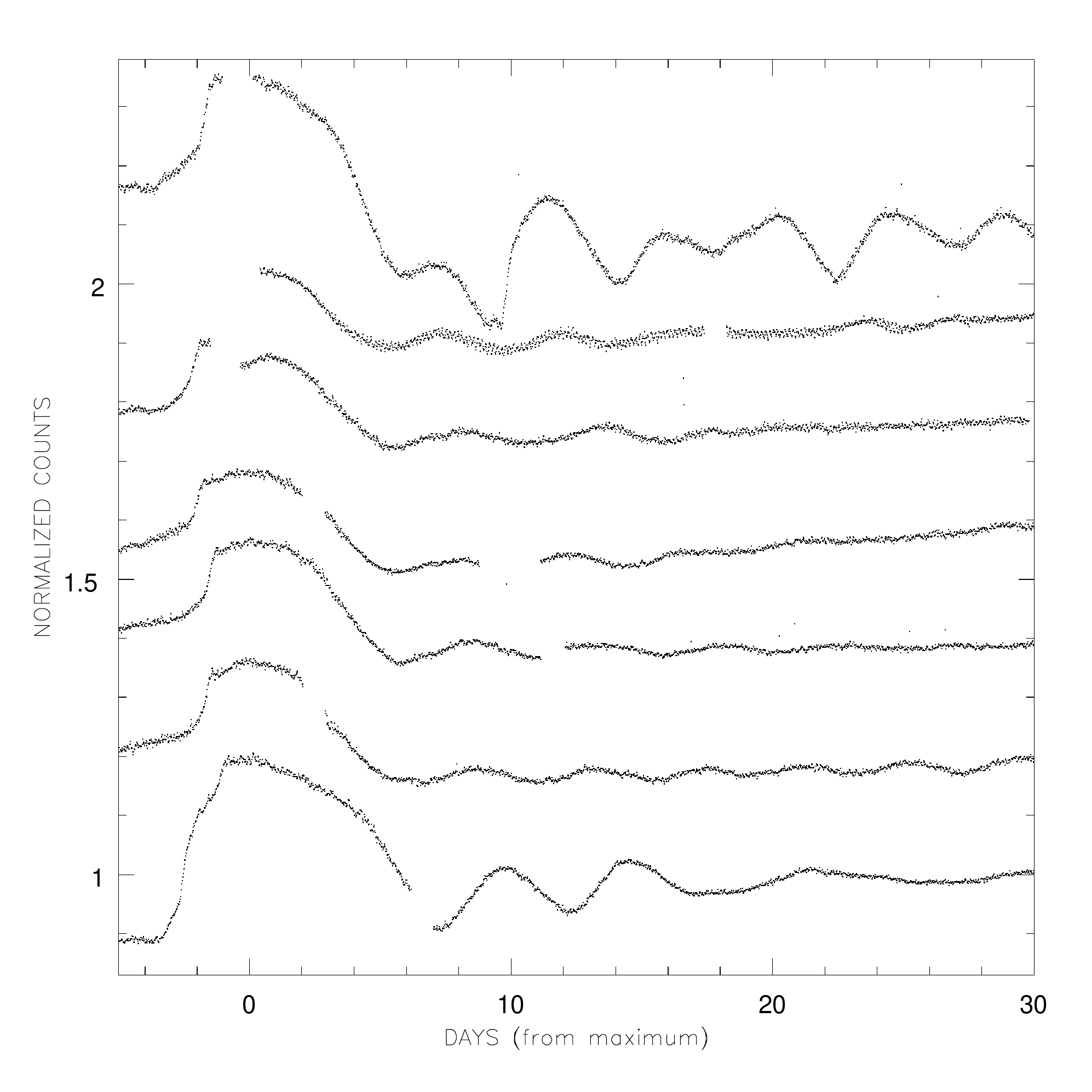}
   \caption{Stack of a selection of ''post outburst'' oscillations observed in V523 Lyr light curve. From bottom to top we plot outbursts 2, X(2), 15, 24, 27 and 28 (see, e.g., Fig.\ref{qrts}). Note that the second light curve from the top clearly shows the small amplitude short period modulation: the 0.151 days period discussed in section 5.2. }  
              \label{dampo}%
   \end{figure}

Last but not least, several of the intra-outburst intervals are characterized by damping oscillations. We sometimes observe just a single one directly following an outburst or see several across the whole quiescence period. These oscillations have varying amplitude, recur on typical time scales of $\sim$3-4 days before fading away and are not strictly periodic (see Fig.\ref{dampo}). Because of this, their period analysis (Sec 5.2) does not provide any reliable result. As much as the oscillations follow an outburst, we believe that the mechanism triggering the oscillations is related to that for the outburst. The latter is still debated (see section 7, Honeycutt 2001, Hameury \& Lasota 2014). However, we imagine that a damping signal might more easily be reproduced/modeled as an accretion disk phenomenon, rather than ascribing it to the secondary or the primary star. 
Similar oscillations are commonly observed among the anomalous Z Cam stars (e.g. V513 Cas, Simonsen 2011) and some of the NL and old-nova systems (e.g. UU Aqr, CP Lac, AH Her, Honeycutt et al. 1998, Honeycutt 2001). 

\subsection{Short and mid-term temporal variations}

\begin{figure}[h!]
   \flushleft
   \includegraphics[width=9cm]{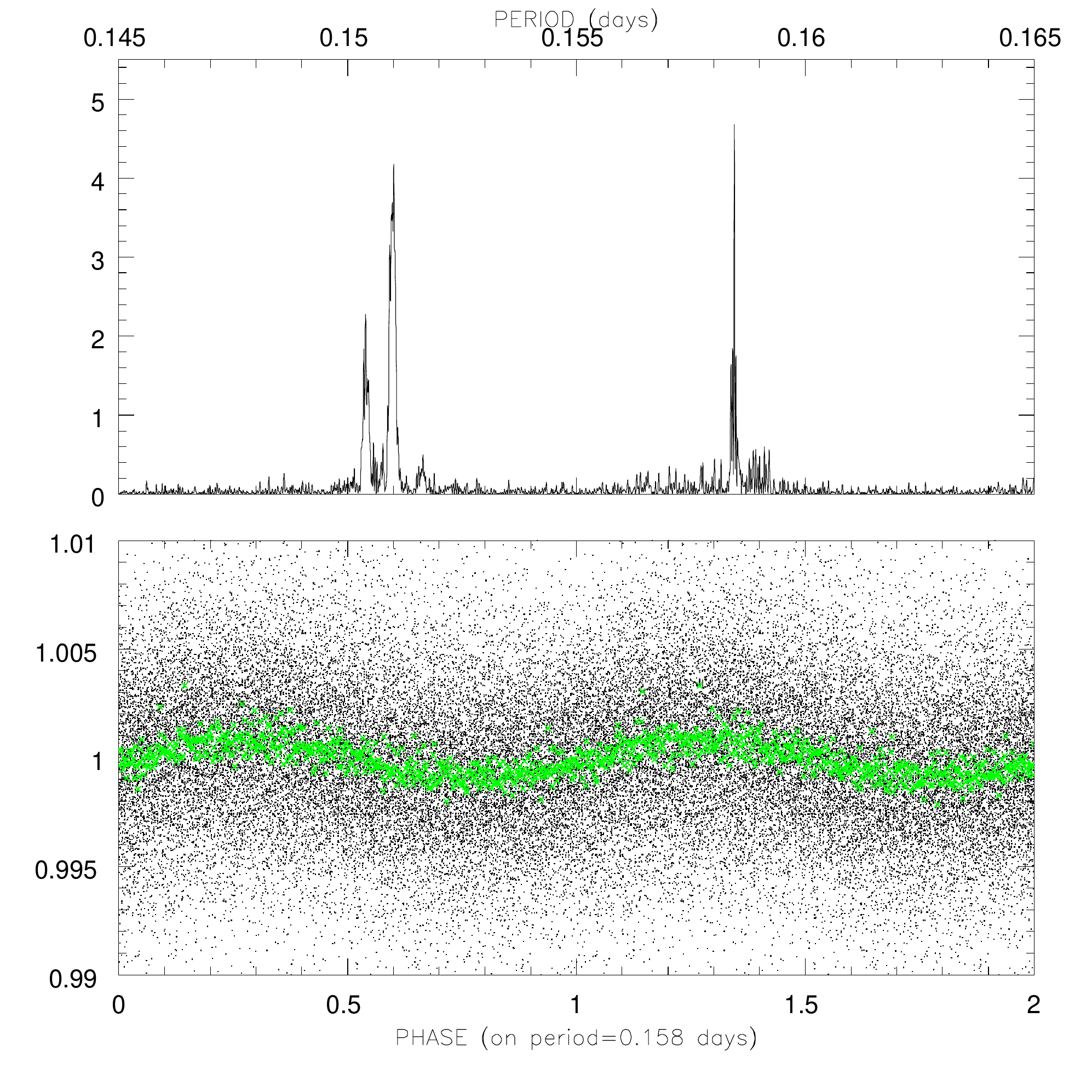}
   \caption{Top panel: periodogram of the 20 quiescent intervals of Fig.3, searched together. Bottom panel: light curve data points from the 20 intervals phased on the period 0.15845058 days ($\sim$3.8 hr, black points). The green crosses represent a phase binned version of the same light curve (bin size 50 data points). }
              \label{pgs}%
    \end{figure}

After the study of the major semi-regular variations such as outbursts and dips, we now move to time-series analysis mainly in search of the orbital period for V523~Lyr. There are many algorithms in the literature to perform period searches. They are not all equivalent as they differ in the ability of handling irregular sampling and/or specific light curve shapes. 
They can be roughly divided in two main groups: the Fourier based and the dispersion/entropy based (see e.g. Engelbrecht 2013 for a quick review). 

Given our past experience with CV light curves and noting other recent works on Kepler CV light curves, we focused our time-series analysis on using the Lomb-Scargle method (LS, Scargle 1982, Horne \& Baliunas 1986) and the phase dispersion minimization technique (PDM, Stellingwerf 1978). The LS periodogram can handle irregular sampling but, similarly to Fourier analysis, is somewhat limited to sinusoidal shape patterns. PDM analysis can also handle irregular sampling but is more sensitive to any repeating pattern and therefore often ideal for non-sinusoidal, irregular light curves.
Fig.\ref{qrts} shows the quiescent intervals that we analyzed first. Pairs of same color vertical segments delimit the interval start and end in each quarter. We explored the period range 0.1--7 days, to include time scales of 3-4 days which are seen to be associated with the damping oscillations. 
This search search did not produce any significant, coherent period of the order of days. However, it revealed the presence of two narrow significant periods at $\sim$0.151 and 0.15845048 days (i.e. 3.6 and 3.8 hr). These are shown in Fig.\ref{pgs}, top panel. The 0.158 day period is extremely narrow in the periodogram and is present through the 20 quiescent intervals marked in Fig.\ref{qrts}. Given its prevalence and extreme coherence we interpret it as the orbital period of V523 Lyr. Note that Garnavich et al. (2015), in a meeting abstract,  report a weak coherent period at 3.8 hr that they too suggest as the orbital period. Fig.\ref{pgs}, bottom panel, shows the quiescent interval data phased on the 0.158 day period, after fit and subtraction of the damping oscillations and trends. The low amplitude of this modulation might be due to the inclination of the binary being nearly face on (see Sect.6). The shorter period is instead split in at least two broader components at 0.1510 and 0.1504 day. These 
periods do not show up in at least the first nine intervals and the majority of their 
power results from intervals 15 through 18. The analysis of these same intervals and their subsets with the PDM technique produced consistent results. 

\begin{figure*}
   \includegraphics[width=20cm,angle=180]{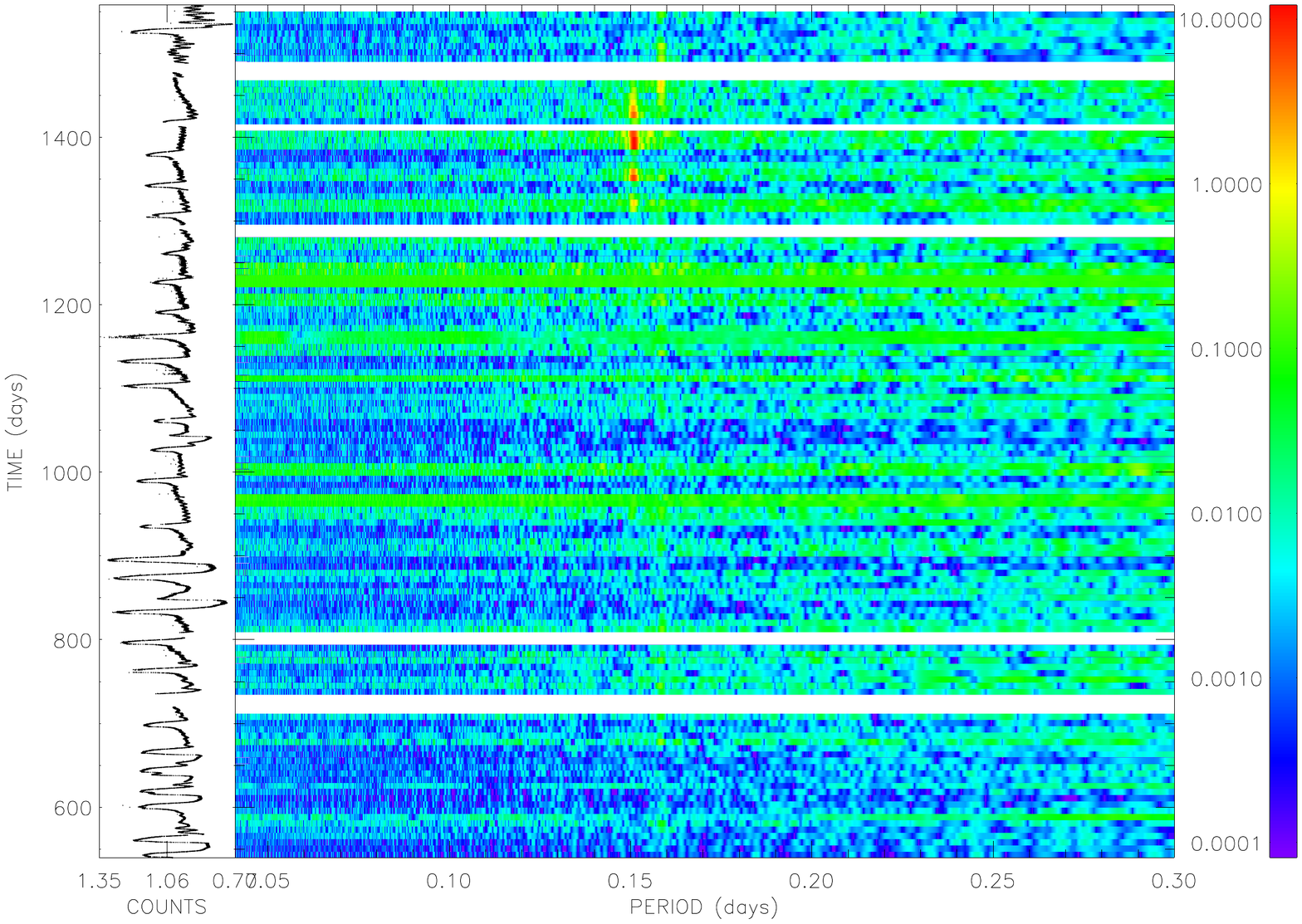}
   \caption{The ``running periodogram'' produced as described in section 5.2. Each 15 days periodogram was normalized to match significance levels thus to be directly comparable to the others. Following the normalization the significance level is at $\geq$1. Note that the color scale is logarithmic (i.e. the whole intensity range encompasses 5 orders of magnitude) in order to make the weak coherent signal at 0.158 days roughly visible.  }
              \label{rs}%
    \end{figure*}

In order to investigate the full light curve (that is including the outbursts), we also created a ``running periodogram'' (Fig.\ref{rs}). We produced periodograms using 15 day subsets of the light curve, offsetting each subset by 7.5 days from the preceding one until we scanned the whole data set. 
The running periodogram analysis shows that the $\sim$0.158 day period is indeed present throughout the entire light curve (the dips possibly being the exception), while the $\sim$0.151 day period is extremely significant only during quarters 14 and 15 (which contain intervals 15 to 18). The 0.151 days period is also clearly visible in the second-from-top light curve shown in Fig. \ref{dampo}. Its value, shorter than that of the orbital period, suggests that it might be interpreted as ``negative superhumps'', i.e. light modulations induced by a tilted (by few degrees) accretion disk whose nodal line is in retrograde precession. The measured period deficit, $\varepsilon_{-}=(P_{-}-P_{orb})/P_{orb}=$-0.047, is consistent to that observed for similarly long orbital period systems (e.g. Table 2 of Wood et al. 2009) and
remarkably similar to the predicted $\varepsilon_-$ of -0.0456 from smoothed particle hydrodynamic (SPH) simulations (Wood et al. 2009, see, in particular their equation {\sc (19)}), adding support the above interpretation for ``negative superhumps''.

\section{Spectroscopic characteristics}
   \begin{figure*}
   \centering
   \includegraphics[width=18cm,angle=0]{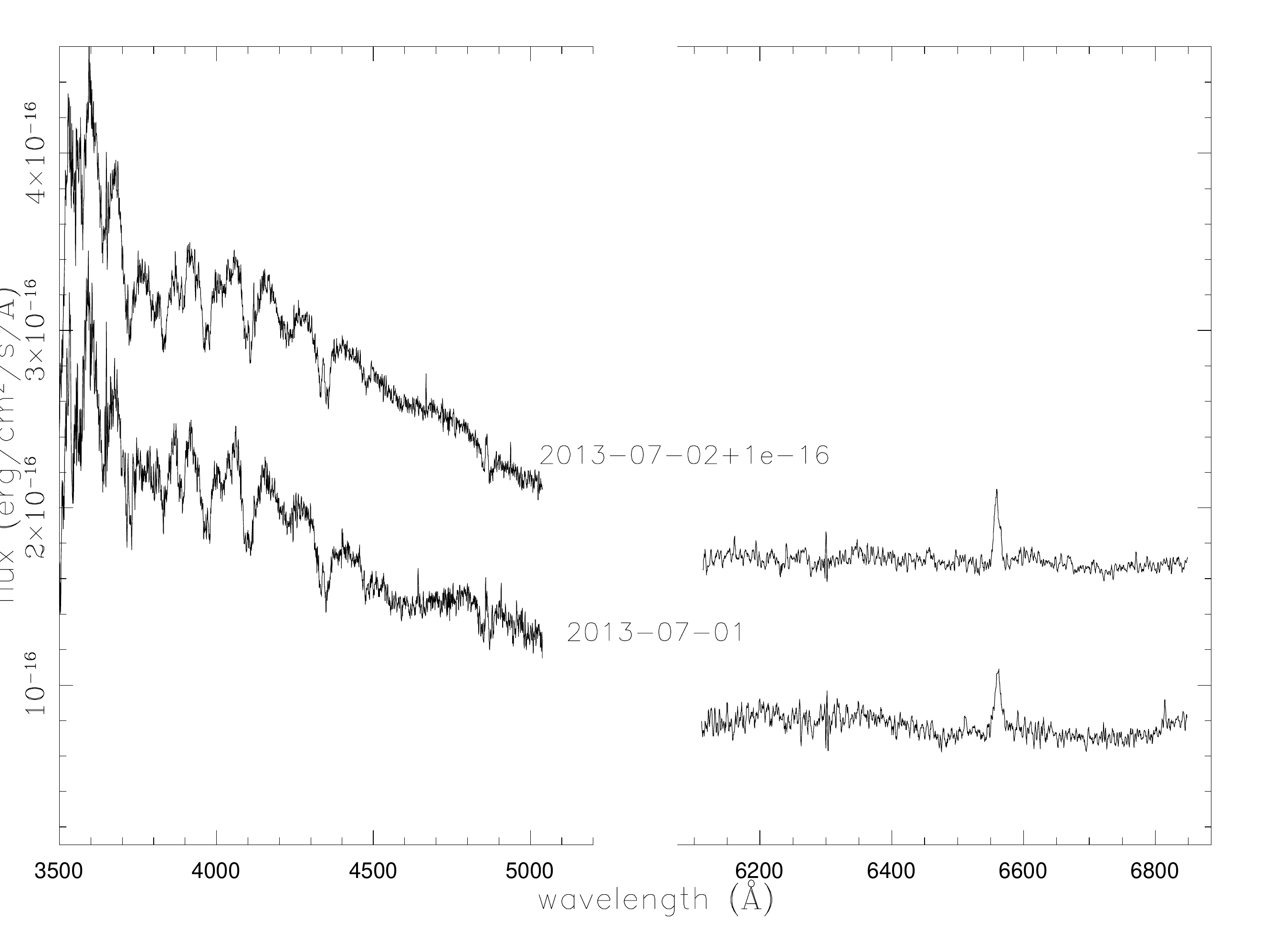}
   \caption{Nightly mean blue and red spectra of V523~Lyr. Note the different total wavelength coverage for each color and the presence of H$\alpha$ in emission while the other Balmer lines are in absorption with emission cores. The night 2 spectra are offset by the amount specified in the figure. The spectra have been smoothed with a running boxcar of size 5. }
              \label{spc}%
    \end{figure*}

Fig.\ref{spc} shows V523 Lyr Hale spectra taken on Jul 1 and 2 2013, a few months after Kepler stopped monitoring the source. Despite collecting several spectra per night, we show just their nightly mean in the figure since the SNR of each individual exposure was low. The red arm exposures has been trimmed (i.e. reduced in spectral range) to show only the region near H$\alpha$ which was placed on a defect free region of the detector.  

During both nights, V523 Lyr shows a blue continuum plus broad H Balmer absorption lines with emission cores; H$\alpha$ is purely in emission. We also see  He\,{\sc i} 4471\AA~in absorption.  We measured FWHM$\sim$500 km/s in the line emission cores of the higher transitions of the Balmer series, H$\alpha$ FWHM being $\sim$600 km/s. 
These spectral characteristics are remarkably similar to those described by Liebert et al. (1993) and Kaluzny et al (1997), although Liebert et al. report a flatter SED, while Kaluzny and collaborators remark that V523 Lyr presented emission cores during the bright/normal state (V=18.13 mag) and no emission cores when $\sim$1 mag fainter. If we consider that V523 Lyr was supposedly $\sim$20 mag in V at the time of the Hale observations, the similarity is puzzling. The fainter spectrum in Kaluzny et al. (1997) was not shown and its resolution is almost a factor of 3 lower than ours, so weak emission cores could have been missed. Contrarily to Liebert et al. and our spectra here, those of Kaluzny et al. do not shown He\,{\sc i} -- which is also puzzling. 

V523 Lyr spectra are very similar to those of the anomalous Z Cam stars V513 Cas and IW And (Szkody et al. 2013). Szkody et al., comparing the outburst and standstill spectra of each object, show that they are very much alike with possibly deeper absorptions and stronger emission cores during outburst and a flatter SED at standstill. It should be noted that IW And when caught in quiescence (i.e. about 2.5 mag fainter than in outburst, based on the flux calibrated spectra presented in Fig.2 of Szkody et al. 2013) shows emission lines. It might be that V523 Lyr was never caught spectroscopically, at any time, in a deep quiescence state: nor at the time of the historic observations nor during our run. 
However, since no precise photometry exists complementing the Hale (and the Kepler) observations, we defer the discussion after further monitoring with simultaneous photometry and spectroscopy has been taken. Regardless of the exact magnitude of V523~Lyr during our observations, they seem to describe a high mass transfer rate CV which is very much alike in spectroscopic appearance and photometric behavior to the Z Cam and (high state) VY Scl stars residing just above the period gap with orbital periods of 3-6 hr.  

While the SNR of the individual spectroscopic exposures was not sufficiently high for separate detailed analysis, we did measure the emission core positions in H$\alpha$ and H$\beta$. The line center shifts between all spectra were found to be $\le$1$\pm$1\AA~(i.e. $\le$62 km/s). For a typical CV with a 3.8 hr period, we'd expect to see radial velocity shifts near $\pm$100 km/sec (or more) for a high orbital inclination system. Therefore the lack of detectable radial velocity motion suggests that V523~Lyr is a nearly face-on binary.

\section{Discussion: V523 Lyr nature}

We have presented the Kepler light curve of V523 Lyr which across $\sim$1000 days shows 28 mini outbursts irregularly spaced (every few to several weeks). Occasionally it shows dips always following an outburst (and never isolated) and/or oscillations. Outbursts and dips are ``mini'' as they appear to be within 0.3 mag amplitude in the Kepler bandpass. 
The Kepler light curve shows all the behaviors observed in the many previous reports, each study observing the system for a limited time. We do not, however, detect any $\sim$3 magnitude deep drops as seen in the past.
Mini-outbursts (of the order of $\Delta m\sim$0.6 mag on average), dips ($\Delta m\sim$0.2-1.0 mag; both following an outburst or isolated) and damping oscillations have been reported by Honeycutt and coworkers (Honeycutt et al. 1998, Honeycutt 2001) for a number of NL and old-novae followed within their RoboScope campaign. They named these mini-outbursts ``stunted-outbursts'' and reported that at least 25\% of NL and old-novae display them. They also claim that the percentage is most likely much higher ($\sim$50\%) if one considers that the majority of NL and old-novae had only sparse monitoring so that the observed variations are not well characterized. 

More recently the subclass of ``anomalous Z Cam stars'' have been identified (Hameury \& Lasota 2014 and references therein). The prototype of such a class is probably RX And whose long term AAVSO light curve has been analyzed by Schreiber et al. (2002), showing that RX And, displaying both Z Cam and VY Scl type behavior (i.e. standstill and time of deep minima), is a transition object between the two classes. Schreiber et al. (2002) describing the light curve write ``RX And shows large variations of its outburst behavior, i.e. periods of frequent short low-amplitude outbursts, irregular inactive states, possibly with mini-outburst activity and typical long dwarf nova outbursts interrupted by {\it disrhythmia}''. 
However, it was only with the detailed temporal monitoring by Simonsen within the AAVSO Z-CamPaign (Simonsen 2011), that ``the bell rang'' and the anomaly of $\Delta m\le$1 mag outbursts were really noticed during the standstill phase of Z Cam stars such as IW And and V513 Cas\footnote{Simonsen (2011) lists five anomalous Z Cam stars but only the two quoted here have good, detailed light curves.} (see also Szkody 2013, Hameury \& Lasota 2014). 

Now, a closer look at the Honeycutt ``stunted outbursters'' and Simonsen's anomalous Z Cam stars reveals that the variability phenomena discussed by those two groups describe essentially identical objects: small outbursts of less that 1 mag in amplitude, occasional short deeps following the small outbursts and low amplitude quasi-periodic oscillations. The only difference is that Honeycutt targets were known NL and old-novae, while Simonsen targets were Z Cam stars. In other words only the latter sample had known, well documented standstills and  low states. If we regard the ``stunted outbursters'' as CVs always in standstill (i.e. always with a mass transfer rate above the critical value -- see e.g. Lasota 2001), we must admit that we are dealing with a relatively homogeneous class of objects and, most important, that we are observing the same astrophysical mechanism, whatever it might be:  whether a variable mass transfer rate (Hameury \& Lasota 2014), or a combination of dwarf nova disk instability and 
localized thermonuclear burning on the surface of the white dwarf (Honeycutt 2001). 
In addition, V523 Lyr's spectral appearance matches the outburst and standstill spectra of the anomalous Z Cam stars IW And and V513 Cas (Szkody et al. 2013), reinforcing the suggested connection between this class of objects and ``stunted outbursters''.  
 
By suggesting that NL, old-nova and anomalous Z Cam stars mini-outbursts rely on the same mechanism, we substantially increase the sample of objects requiring the same astrophysical explanation. We notice that in addition to the objects targeted by the RoboScope and Z-CamPaign, a significant fraction of the NL monitored by Kepler are being classified as ``stunted outbursters'' (see, e.g. Gies et al. 2013, Ramsay et al. 2016). Hence, 
this type of object appears to be common, therefore the mechanisms involved to explain their behavior need to be reconsidered and elucidated. 

\section{Summary and conclusion}

To conclude, the results of the presented analysis can be summarized as follow: 

\noindent \_ V523 Lyr Kepler light curve shows numerous mini-outbursts of amplitude $\leq$0.3 mag in the Kepler bandpass, with durations of $\sim$1 week and recurring with an average time scale of 33$\pm$9.4 days. The mini-outbursts are often followed by dips, $\sim$0.1-0.2 Kepler-magnitudes deep and lasting about 1 week and sometimes by damping oscillations as well. The time series analysis of the light curve shows the presence of an low amplitude persistent and coherent period of 0.15845058~day ($\simeq$3.8~hr) which we identify with the orbital period of the system. The analysis also shows the presence of a strong temporary signal at $\sim$0.151 days which is consistent with nodal or negative superhumps.

\noindent \_ The Hale spectra show that V523 Lyr has a blue SED with broad H Balmer absorption lines containing emission cores. The spectra did not show intra- or across-night radial velocity variations suggesting a quite low orbital inclination. 

\noindent \_ The V523 Lyr Kepler light curve and the light curves of the ``stunted outbursters'' described by Honeycutt et al. are a close match. In addition, when combined with  the historic photometric monitoring in the literature we see that V523~Lyr is an anomalous Z-Cam star. Its spectra support the same conclusion. Furthermore, we suggest that anomalous Z Cam and ``stunted outbursters'' mini-outbursts are produced by the same mechanism. We also suggest that Honeycutt's ``stunted outbursters'' are systems always in standstill, i.e. with a mass transfer rate above the critical value for disk stability. 

\begin{appendix}
\section{Variable stars in V523 Lyr Kepler aperture}

   \begin{figure*}
   \flushleft
   \includegraphics[height=17cm,angle=270]{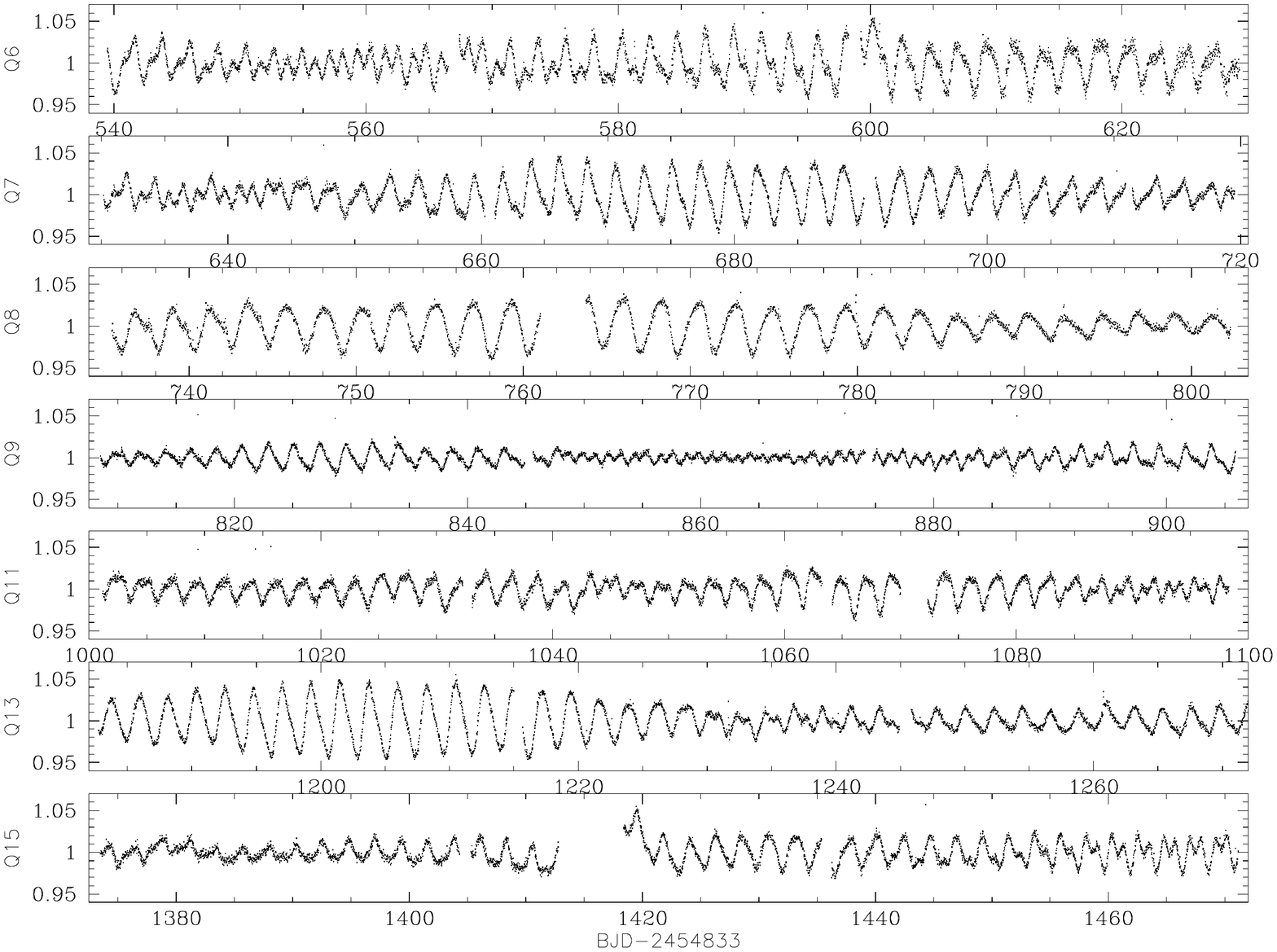}
   \includegraphics[height=17cm,angle=270]{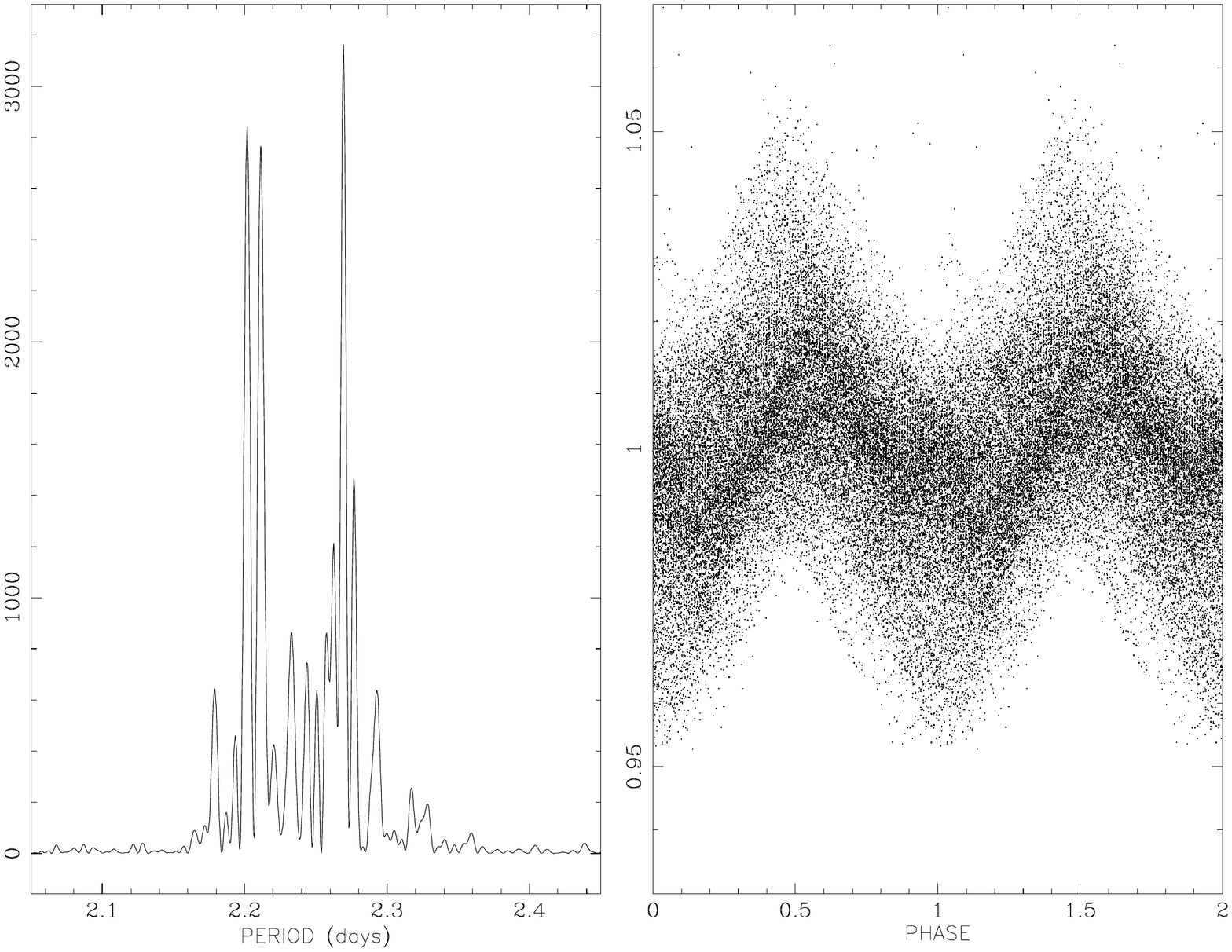}
   \caption{Top panel: Light curve of the variable S2. Bottom panel: LS periodogram (left) and light curve phased on the most significant period (2.2694 days).}
              \label{s2}%
    \end{figure*}

As mentioned in Sect.3, when examining the V523 Lyr apertures on a pixel-by-pixel basis, we found two other sources which are
variable and potentially interesting. Neither was present as a contaminating source for V523~Lyr in every quarter since their inclusion in the SAP light curve depends on the Kepler imager orientation and the SAP aperture size. We summarize below observational details and period analysis for these two variables which we called S2 and S3 (V523 Lyr, being implicitly ``star~1''/S1).

S2 is located $\sim$2-2.5 pixels W and 2.5-3 pixels S of V523 Lyr and could match one of the two stars circled in blue in 
Fig.\ref{fov}. It is present in the V523~Lyr SAP light curve during quarters 6,7,8,9,11,13 and 15. 
S3 is about 6.5 pixels E and 4 pixels N of V523 Lyr. Its probable identification has been marked with the label S3 in Fig.\ref{fov}. Should S3 match the object indicated in the finding chart, it corresponds to a red star of B=18.93, R=18.32 and I=17.88 mag, according to the USNO-B1 catalog. S3 is present in the V523~Lyr SAP light curve during quarter 6 and 7, in the latter case only the wings of its PSF are included in the aperture. We remark that for both of these variables, the distance from V523 Lyr and their listed identifications are uncertain\footnote{Without a reliable PSF for centroiding, positional error might be as large as 0.5-1.0 Kepler pixels.} and should be taken with extreme caution.

\begin{table}
 \label{}
 \caption{Apertures used herein for the variable stars S2 and S3, when present in the V523~Lyr SAP target pixels. Pixel coordinates (x,y) correspond to the CCD columns and raws (see header keywords {\sc crval1p} and {\sc crval2p}). }
\centering
\begin{tabular}{ccl}
QRT & SIZE & PIX COORDs (COL,RAW) \\ \hline
\multicolumn{3}{c}{ star S2} \\ \hline
6 & 1 & (164,848) \\ 
7 & 3 & (163,843) (164,842) (164,843) \\ 
8 & 3 & (158,842) (159,842) (159,843) \\ 
9 & 1 & (159,849) \\ 
11 & 3 & (163,843) (163,842) (164,843) \\ 
13 & 2 & (159,849) (159,848) \\ 
15 & 3 & (163,843) (163,842) (164,843) \\ \hline 
\multicolumn{3}{c}{ star S3} \\ \hline
6 & 3 & (155,856) (155,855) (156,856)  \\ 
\end{tabular}
\end{table}

The pixels extracted for each of the confounding stars (in the quarters they were present in the V523~Lyr aperture) are listed in Table A.1. 
Both stars were easily noticed for they show a comparatively large amplitude, periodic pattern. After extraction of their light curves, we performed a period analysis for both. The light curves for the variable stars S2 and S3 
were ``detrended'' before performing any period analysis. The detrending was done simply by dividing each quarter's light curve by its heavily smoothed version, i.e., using an extremely large running box (namely $\sim$500 data points corresponding to $\sim$10 days). This ensures that the inherent oscillations remain unchanged as they are divided by a local ``continuum''. 

The light curve of S2 is shown in the top panel of Fig.\ref{s2}. The light curves appears to be caused by a rotational modulation with varying star spot amplitudes. The LS analysis of the complete light curve produces the periodogram shown in the bottom left panel of Fig.\ref{s2}; while in the right panel of the same figure we show the light curve phased on the most significant period (i.e. 2.2694 days).



\begin{figure}
   \flushleft
   \includegraphics[width=7.5cm,angle=270]{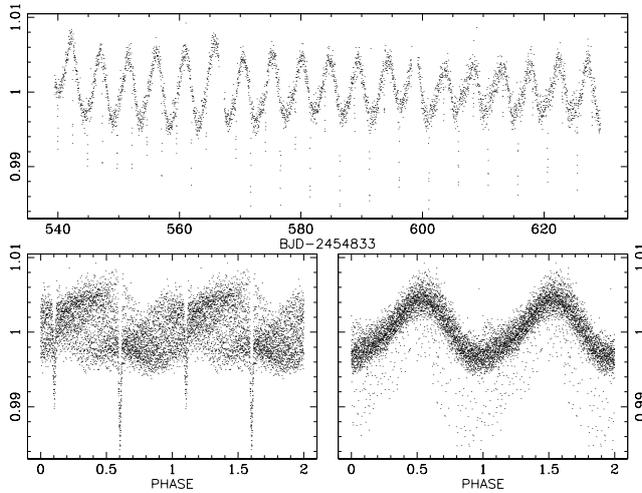}
   \caption{S3 light curves. Top panel: the normalized light curve extracted from quarter 6. Bottom panels: the same light curve phased on the period 2.443098 days, clearly showing that the true eclipse period is twice this value (left) and the best fit period for the sine-like oscillations, P= 4.742325 days (right).}
              \label{s3}%
    \end{figure}

Fig.\ref{s3} top panel shows the light curve extracted for S3 from quarter 6 observations. S3 shows alternating deep and shallow eclipses and a strongly sinusoidal modulation. The light curve was analyzed with  the box-fitting least squares (BLS) algorithm (most appropriate for eclipse searches, Kovacs et al., 2002) yielding the eclipse period (primary to primary eclipse) of 4.881696 days; while the best fit sinusoid period is 4.742325 days. We note that the sinusoidal modulation cannot be due to ellipsoidal variations since the period does not match that of the eclipses and their relative phase changes constantly. Perhaps this indicates that one of the stars pulsates or that there is a rotational modulation not yet tidally locked to the binary period. 

\end{appendix}
    
\begin{acknowledgements}
The NASA Kepler Mission was selected as the 10th mission of the Discovery Program. Funding for this mission is provided by NASA's Science Mission Directorate. This research has made use of the NASA Exoplanet Archive, which is operated by the California Institute of Technology, under contract with the National Aeronautics and Space Administration under the Exoplanet Exploration Program. SH wishes to thank the staff at the Mt. Palomar Hale 200-inch telescope for their help and expertise during the collection of the spectra and Sally Seebode and Dawn Gelino for help with the observations. EM thanks Steven Shore for the confrontations as well as the occasionally strong criticisms.  EM wishes to thank also Carlo Morossi for being the ideal ``office-mate'' always ready to help and take questions and the Kepler GO team for the support with PyKE. The authors thank the anonymous referee for the careful reading of the manuscript and the valuable suggestions. 
\end{acknowledgements}

\end{document}